\documentstyle[aps,floats,12pt]{revtex}
\tighten
\begin{document}
\draft
%
%
\title{Molecular-Dynamics Simulation of a Glassy Polymer Melt:\\
Incoherent Scattering Function}
\author{C. Bennemann, J. Baschnagel\footnote{To whom correspondence should be addressed.  
Email: {\sf baschnag@flory.physik.uni-mainz.de}}, and W. Paul\\[2mm]}
\address{Institut f\"ur Physik, Johannes-Gutenberg Universit\"at,\\Staudinger
Weg 7, D-55099 Mainz, Germany}
\maketitle
%
%
%
%
\newcommand{\mr}[1]{{\rm #1}}
\renewcommand{\vec}[1]{\mbox{\boldmath$#1$\unboldmath}}
%
%
\input{epsf}
%
%
%
%
\bigskip
\begin{center}
{\bf Abstract}
\end{center}
\begin{abstract}
We present simulation results for a model polymer melt, consisting of
short, nonentangled chains, in the supercooled state. The analysis 
focuses on the monomer dynamics, which is monitored by the incoherent
intermediate scattering function. The scattering function is recorded over 
six decades in time and for many different wave-vectors which range from the
size of a chain to about three times the maximum position of the static
structure factor. The lowest temperatures studied are slightly above 
$T_\mr{c}$, the critical temperature of mode-coupling theory (MCT), 
where $T_\mr{c}$ was determined from a
quantitative analysis of the $\beta$- and $\alpha$-relaxations. 
We find evidence 
for the space-time factorization theorem in the $\beta$-relaxation regime, and for 
the time-temperature superposition principle in the $\alpha$-regime, if the 
temperature is not too close to $T_\mr{c}$. The wave-vector ($q$-) dependence of 
the nonergodicity parameter, of the critical amplitude, and the $\alpha$-relaxation
time are in qualitative agreement with calculations for hard spheres. For $q$
larger than the maximum of the structure factor the $\alpha$-relaxation time $\tau_q$ 
already agrees fairly well with the asymptotic MCT-prediction $\tau_q \sim q^{-1/b}$. 
The behavior of the 
relaxation time at small $q$ can be rationalized by the validity of the 
Gaussian approximation and the value of the Kohlrausch stretching exponent, as
suggested in neutron-scattering experiments.
\end{abstract}
%
%
\pacs{{\sf PACS}: 61.20.Ja,64.70.Pf,61.25.Hq,83.10.Nn\\
submitted to {\em Phys.\ Rev.\ E} on \today}
%
%
\section{Introduction}
\label{intro}
During the past decade, numerous experiments and simulations have focused
attention on the dynamics of supercooled liquids in a temperature region
about fifty degrees above the calorimetric glass transition
\cite{ali_93,yip,kob_rev97,kob_rev95,goetze2}. This interest
has been elicited by the development of the so-called mode-coupling theory
(MCT) \cite{goetze2,goetze1,goetze_yip}. Mode-coupling theory predicts 
that there is a critical temperature
$T_\mr{c}$ above $T_\mr{g}$, where the dynamics of the glass former
qualitatively changes from a liquid-like to a solid-like behavior. The onset of
this change is manifested by a two-step decay of dynamic correlation functions,
which couple to density fluctuations. These correlation functions can be
measured in experiments and computer simulations. The distinguishing feature of the
theory is that it makes universal, system-independent predictions about 
the shape of the time correlation functions, that there are constraining
relationships between various theoretical quantities, and that key,
nonuniversal parameters can be expressed in terms of the glass former's static
structure.

Especially the latter point opens the possibility of a direct
comparison between theory and experiment or simulation. Such comparisons are,
however, predicated upon having accurate data for the temperature dependence of
the static structure factor at hand. Therefore, they have only been
performed for few,
simple systems, such as hard-sphere-like colloidal particles \cite{gs_97} or computer
models of soft spheres \cite{bl}, and binary Lennard-Jones mixtures \cite{nk}.

On the other hand, most tests of MCT have concentrated on the
universal predictions by adjusting the system-specific parameters. This fit
procedure must simultaneously optimize at least three free parameters. It is
rather involved and further complicated by the fact that the theoretical
predictions are only valid asymptotically in a narrow temperature interval
around $T_\mr{c}$, and that the microscopic
(vibrational) time scales have to be well separated from those of the structural
relaxation. These drawbacks have led to a criticism of the significance of the
experimental evidence for the theory \cite{ki-cu}, but also to extensions of MCT to 
tackle problems, like the interference of vibrational and relaxational time scales
\cite{tgms}, orientational degrees of freedom \cite{ss}, and corrections to the 
asymptotic behavior \cite{fgm,ffgms}.
This situation suggests that further tests may be beneficial.

In the present paper we want to discuss the results of a molecular-dynamics (MD)
simulation for a polymer melt and the analysis of the incoherent intermediate
scattering function by the idealized mode-coupling theory. Comparable
applications of the theory to polymer melts have already been performed in
experiments and simulations. One of the earliest applications were neutron
scattering experiments of polybutadiene \cite{zrff}. These experiments found the 
predicted square-root behavior for the temperature dependence of the nonergodicity
parameter below $T_\mr{c}$, suggested an in-phase variation of the coherent
$\alpha$-relaxation time with the static structure factor, and provided
evidence for time-temperature superposition above $T_\mr{c}$.
These studies mainly 
focused on the $\alpha$-relaxation. A detailed line shape analysis in the 
$\beta$-relaxation regime was not made. Such an analysis was
attempted in dielectric relaxation experiments of poly(ethylene terephthalate)
\cite{hkf}. The experiments are complicated by the fact that poly(ethylene 
terephthalate) has a high tendency to crystallize, and that a $\beta$-peak masks 
the MCT $\beta$-process. Nevertheless, the experiments confirmed some of the 
idealized predictions, but also claimed that there are severe deviations,
especially below $T_\mr{c}$. Close to and below $T_\mr{c}$ deviations are expected 
due to ergodicity restoring processes. Usually, the extended MCT \cite{fghl}
is then applied to account for these processes. However, a recent reanalysis 
\cite{ems} of the dielectric data indicates that the deviations cannot be explained 
in this way. Instead, a higher order scenario, an $A_3$-singularity, has to be used. 
In applications to other structural glass formers these higher-order singularities 
are less common, but some studies suggest that they could be pertinent to 
partially crystalline polymers \cite{sj_1991,ion}.

On the simulation side, the glassy behavior of polymer melts has been studied
by Monte-Carlo simulations of the bond-fluctuation lattice model
\cite{bbbp,rev96,kb_rev93}. These simulations deal with short, nonentangled chains, 
and have been restricted to temperatures above $T_\mr{c}$. A comparison with MCT 
indicated that a quantitative description of the $\beta$-relaxation is possible if 
the extended theory is used \cite{exmct}. Despite this agreement, there are still 
some points which could be improved upon. On the one hand, the 
simulations were performed at constant volume, whereas experiments are usually done at
constant pressure, and the underlying lattice structure precludes all phonon
contributions to the short-time dynamics. On the other hand, the simulation
data were not completely equilibrated. They still exhibited very slow physical
aging processes. These aging processes effectively correspond to additional relaxation 
channels which are not contained in the idealized MCT.
Although it is interesting that the extended MCT is able
to describe such a situation -- especially when taking into account the current
intensive research on physical aging below $T_\mr{c}$ \cite{kb,bckm} --, the 
simulations do not meet the theoretical premise of thermal equilibrium.

Both drawbacks are removed by the present MD-simulation. It is done at constant
pressure and equilibrated on all length scales which we will discuss, i.e., at
low temperatures the simulation time spent for equilibration exceeded by one 
order of magnitude the time,
over which actual dynamical measurements were performed after equilibration. 
Some details of the 
model, of its static properties, and of the simulation are compiled in the next section. 
The subsequent section is split into two parts. The first describes the analysis of
the $\beta$-relaxation, and the second that of the $\alpha$-relaxation. The
last section summarizes the main results.
\section{Model, Static Results, and Simulation Technique}
\label{modstasim}
This section briefly reviews some characteristics of the model. A detailed description 
of its properties and of the simulation technique can be found in Ref.~\cite{bpbd}.

The simulations were done with linear, nonentangled, monodisperse chains of
length $N=10$. All monomers interact by a truncated and shifted Lennard-Jones
(LJ) potential, $U_{\mr{LJ}}(r)=4\epsilon[(\sigma/r)^{12}-(\sigma/r)^6] + C$.
The constant $C=0.00775$ assures that the potential 
vanishes if $r \geq 2r_\mr{min}=2\times 2^{1/6} \sigma$, where $r_\mr{min}$ is the 
minimum position.
Temperature and distances are measured in units of $\epsilon/k_{\mr{B}}$ and $\sigma$, 
respectively, and time is measured in units of $(m\sigma^2/\epsilon)^{1/2}$, 
where the mass is set to unity.

In addition, there is a FENE-potential between bonded monomers
along the backbone of a chain, i.e., $U_\mr{F}(r)=-15R_0^2\ln[1-
(r/R_0)^2]$ with $R_0=1.5$. With these parameters the superposition of
the FENE- and LJ-potentials generates a steep effective bond-potential with
a minimum at about $0.96\sigma$ 

In our study a melt configuration contained 120 polymers, and ten configurations
were simulated at each temperature to improve the statistics. The simulation procedure
consisted of two steps. First, the volume was allowed to fluctuate in a simulation
at a given temperature and pressure (always $p=1$ -- the 
influence of pressure is studied in Ref.~\cite{bpbb_pressure}) to determine the
equilibrium density at this thermodynamic state point.
Fixing then the resulting volume the subsequent runs used the Nos\'e-Hoover
thermostat to simulate in the canonical ensemble. All dynamic properties calculated
are results from these canonical simulations. In order to equilibrate the system, 
each chain was propagated several times over the distance of radius of gyration before 
starting the analysis. This time suffices so that the incoherent scattering functions
have decayed to zero for all wave-vectors (except for $q < 2$ at $T=0.46$). 

The particular choice of the bonded and nonbonded potentials has two main consequences 
for the static properties of the model. First, the chains do not become stiffer with decreasing
temperature. In the interesting temperature region ($T < 0.7$) the end-to-end
distance, $R_\mr{e}$, and the radius of gyration, $R_\mr{g}$, are essentially
constant: $R_\mr{e}^2=12.3\pm 0.1$, $R^2_\mr{g}=2.09\pm 0.01$. Second, the monomer
distance 0.96, favored by the bond-potential, is incompatible with 
multiples of $r_\mr{min}$, and thus with crystalline ordering. This is 
illustrated by the static structure factor of the melt in 
\begin{figure}
\begin{center}
\begin{minipage}[t]{100mm}
\epsfysize=90mm
\epsffile{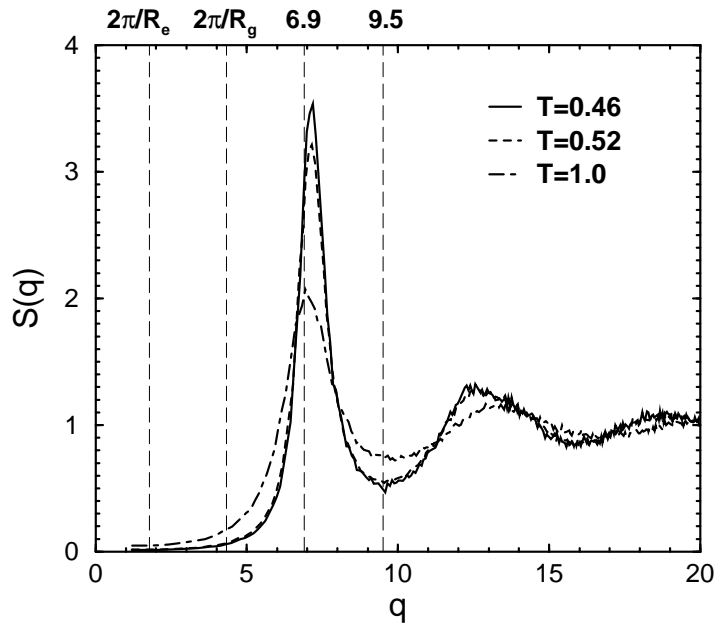}
\end{minipage}
\end{center}
\caption[]{Temperature dependence of the melt's structure factor. The temperatures
span the interval from the normal liquid ($T=1$) to the supercooled state ($T=0.46$)
of the melt. The lowest temperature is slightly above $T_\mr{c} \simeq 0.45$. 
In the mode-coupling analysis mostly $q=3,6.9,9.5$ are used. The smallest $q$-value
probes the size of a chain (end-to-end distance: $R_\mr{e}\simeq 12.3$, radius of gyration:
$R_\mr{g}\simeq 2.09$), whereas the larger wave-vectors correspond to intermonomer 
distances.}
\label{q_joerg}
\end{figure}
Fig.~\ref{q_joerg}\footnote{This figure extends the data of Fig.~6 in 
Ref.~\cite{bpbd}. Note
that we chose the conventional normalization in the present case contrary to
Ref.~\cite{bpbd}. In Fig.~\ref{q_joerg}, $S(q)$ tends to the
isothermal compressibilty divided by that of the ideal gas, {i.e.}, to
$\rho T \kappa_T$, if $q \rightarrow 0$, whereas it approaches $(\rho T \kappa_T/
\mbox{total number of monomers})$ in Ref.~\cite{bpbd}.}. At small wave-vectors, the
structure factor is of the order of 0.01, indicating that the melt has a low
compressibility, then it raises and exhibits an amorphous halo close to $q=6.9$,
corresponding to the nearest-neighbor packing of the monomers, and finally
decreases again and begins oscillating to gradually approach 1 for large $q$. 
This structure is characteristic of the liquid state and present at all 
temperatures. With decreasing temperature the maxima and minima become
sharper, and the position of the first maximum slightly shifts to larger 
$q$-values because the density of the melt increases.
\section{Idealized Mode-Coupling Analysis}
Close to the critical temperature mode-coupling theory predicts a two-step
relaxation behavior for dynamic correlation functions that couple to density
fluctuations, like the incoherent intermediate scattering function,
\begin{equation}
\phi_q^\mr{s}(t)=\frac{1}{M} \sum_{m=1}^M \bigg \langle
\exp\left(\mr{i}\vec{q}\cdot [\vec{r}_m(t)-\vec{r}_m(0)]\right)
\bigg \rangle \; ,
\label{eq1}
\end{equation}
where $M$ stands for the total number of monomers in the melt. The expected
two-step relaxation gradually develops for $T\leq 0.55$ in our model. Therefore
the following analysis focuses on this temperature region.
It is divided into two parts. First, we discuss the
$\beta$-relaxation regime, and then the final structural decay, the
$\alpha$-relaxation.
\subsection{$\beta$-relaxation regime}
\label{betrel}
The $\beta$-relaxation regime is defined as the time window, where the
correlator $\phi_q^\mr{s}(t)$ is close to the 
nonergodicity parameter $f_q^\mr{sc}$, {i.e.}, $|\phi_q^\mr{s}(t)-
f_q^\mr{sc}| \ll 1$ \cite{goetze2,goetze1,goetze_yip,fgm,ffgms}. The corrections to $f_q^\mr{sc}$ 
take the following form \cite{fgm}
\begin{equation}
\phi_q^\mr{s}(t)
 =  f_q^\mr{sc}+h_q^\mr{s}G(t)
 +  h_q^{\mr{s}(2)}\left[\frac{t}{\tau}\right]^{2b} \; .
\label{eq2}
\end{equation}
The first two terms represent the space-time factorization theorem. The spatial
dependence is contained in $f_q^\mr{sc}$ and in the 
critical amplitude $h_q^\mr{s}$, whereas the $\beta$-correlator $G(t)$ carries
the whole time and temperature dependence. In the idealized MCT it is given
by \cite{goetze1,fgm,ffgms}
\begin{equation}
G(t)=\sqrt{|\varepsilon|} g(t/t_\varepsilon)
\quad \mbox{with} \quad t_\varepsilon=\frac{t_0}{|\varepsilon|^{1/2a}}\, ,
\; \; \varepsilon=C\left [\frac{T_\mr{c}-T}{T_\mr{c}} \right ]\; ,
\label{eq3}
\end{equation}
where $t_0$ is a matching time to the microscopic transient, $t_\varepsilon$
is the $\beta$-time scale, $\varepsilon$ is the separation parameter ($C=$ 
constant), and $g(\hat{t})$ is the temperature independent
$\beta$ master function. Its shape is determined by the 
exponent parameter $\lambda$, which in turn fixes the critical
exponent $a$ and the von-Schweidler exponent $b$ via
$\lambda=\Gamma(1-a)^2/\Gamma(1-2a)=\Gamma(1+b)^2/\Gamma(1+2b)$.

The third term of Eq.~(\ref{eq2}) is a leading-order
correction (of order $|\varepsilon|$) to $h_q^\mr{s}G(t)$ (which is of
order $|\varepsilon|^{1/2}$) \cite{fgm,ffgms}. 
It extends the description of the decay from $f_q^\mr{sc}$ to longer times ({i.e.}, 
for $t_\varepsilon \ll t \ll \tau$ it is a correction to the von-Schweidler law $t^b$),
and depends on temperature by the $\alpha$-time scale $\tau$
\begin{equation}
\tau=t_\varepsilon\left [\frac{t_\varepsilon}{t_0}\right ]^{a/b}
=\frac{t_0}{|\varepsilon|^\gamma} \quad \mbox{for} \quad T \geq T_\mr{c}
\label{eq4}
\end{equation}
with $\gamma=1/(2a)+1/(2b)$.

When applying these formulas to the simulation data we proceeded in two
steps. First, we tried to fix the exponent parameter by working with
the factorization theorem only. To this end, we calculated the $\beta$-scaling 
function for a specific value of $\lambda$ numerically. The result was
inserted in Eq.~(\ref{eq2}), and the remaining parameter, $f_q^\mr{sc}$,
$t_\varepsilon$, and the total prefactor of $g(\hat{t})$, $\tilde{h}_q^\mr{s}(T)
=h_q^\mr{s} |\varepsilon|^{1/2}$, were adjusted at a given temperature and $q$-value. This 
procedure was repeated for different $\lambda$- and $q$-values (mainly 
$q=6.9, 9.5$), and at several temperatures (mainly $T=0.48,0.52$) to explore
which range of $\lambda$-values yielded fits of comparable quality. 
The best-fit result 
was:
\begin{equation}
\lambda=0.635\pm 0.025\;, \;\; a=0.352\pm 0.010\;, \;\; b=0.75\pm 0.04\;, \; \;
\gamma=2.09\pm 0.07 \; .
\label{mct_para}
\end{equation}
In addition, this first step also helps to find the upper bound of the temperature
interval, where the asymptotic formulas of MCT can be applied. For the model
studied this is $T \approx 0.52$. Therefore, the second step was restricted
to $T \leq 0.52$. In this step, we fixed $\lambda=0.635$ and optimized
the remaining parameters, $f_q^\mr{sc}$, $t_\varepsilon$, $\tilde{h}_q^\mr{s}$, and
$\tilde{h}_q^{\mr{s}(2)}(T)=h_q^{\mr{s}(2)} \tau^{-2b}$. This procedure was done for 
$T=0.47,0.48, 0.49,0.5,0.52$, and at about 20 different $q$-values, ranging from $q=1$ 
to $q=19$, for each temperature.

\begin{figure}
\begin{center}
\begin{minipage}[t]{100mm}
\epsfysize=90mm
\epsffile{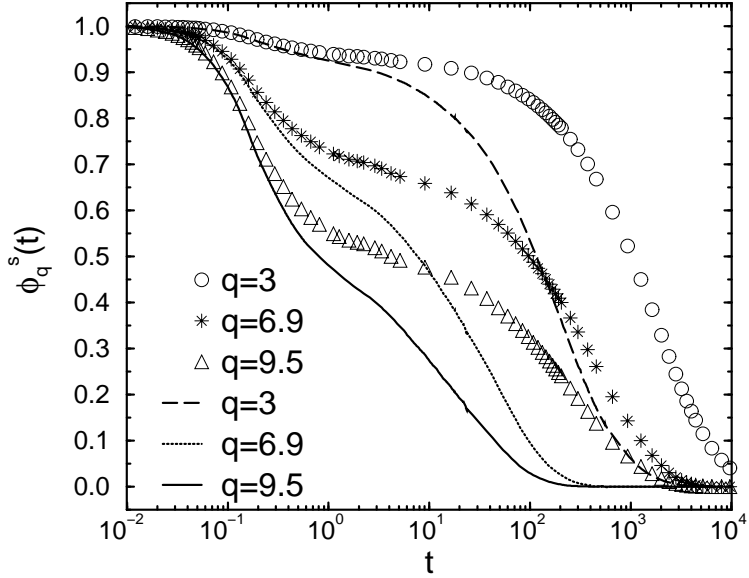}
\end{minipage}
\end{center}
\caption[]{Incoherent intermediate scattering functions versus time 
at $T=0.47$ (symbols) and at $T=0.52$ (lines). Three
$q$-values are shown: $q=3$ ($\approx$ size of the chain), $q=6.9$ [$\approx$ maximum of
$S(q)$], and $q=9.5$ [$\approx$ first minimum of $S(q)$].}
\label{q=all.T=0.47+0.52}
\end{figure}

The simulation results for $T=0.47$ and $T=0.52$ are 
compared in Fig.~\ref{q=all.T=0.47+0.52}.
The figure shows the incoherent intermediate scattering 
function over six decades in time\footnote{Time is measured in units of the MD time step,
which is $\mr{d}t=0.002$ \cite{bpbd}. Six decades in time 
therefore correspond to $5\times 10^6$ MD steps.}
for three different wave vectors.
The smallest $q$-value ($q=3$) probes the length scale of a
chain, whereas $q=6.9,9.5$ approximately correspond to the 
maximum and the first minimum of the structure factor $S(q)$ 
(see Fig.~\ref{q_joerg}). At a given $q$-value the simulation 
data almost coincide for both temperatures if $t \leq
10^{-1}$. This short-time regime corresponds to the ballistic
motion of a monomer, which only shows a weak $\sqrt{T}$  temperature
dependence. Contrary to that, the curves strongly separate from 
one another with increasing time. At $T=0.47$, $\phi_q^\mr{s}(t)$
reaches zero about a decade later than at
$T=0.52$. To achieve a similar growth of the structural
relaxation time when cooling from higher temperature to
$T=0.52$, one has to take $T\approx 0.7$, a temperature that
is $35\%$ larger than $T=0.52$. Compared to this
difference, the $10\%$ disparity between $T=0.47$
and $T=0.52$ is indicative of the approach to $T_\mr{c}$,
where the asymptotic formulas, Eqs.~(\ref{eq2})--(\ref{eq4}),
should hold.

\begin{figure}
\begin{center}
\begin{minipage}[t]{100mm}
\epsfysize=90mm
\epsffile{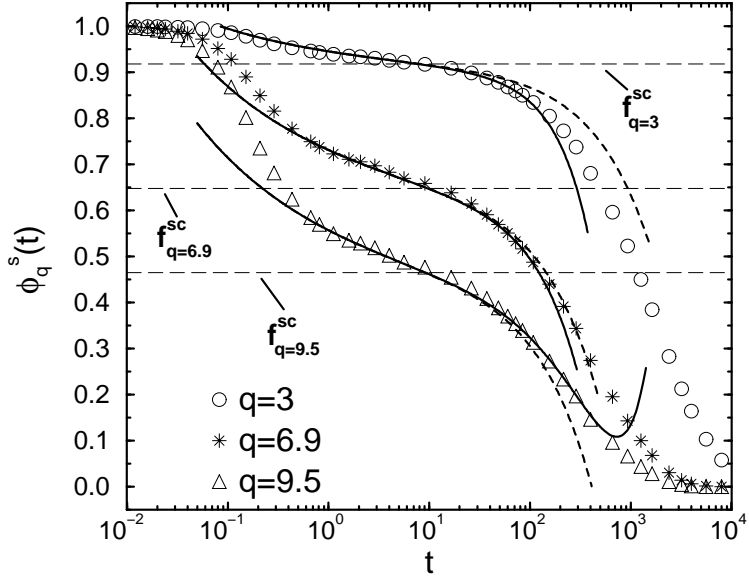}
\end{minipage}
\end{center}
\caption[]{Incoherent intermediate scattering function versus time at $T=0.47$. Three
$q$-values are shown: $q=3$ ($\approx$ size of the chain), $q=6.9$ [$\approx$ maximum of
$S(q)$], and $q=9.5$ [$\approx$ first minimum of $S(q)$]. The dashed lines are the 
MCT-fit results, using only the factorization theorem, whereas the solid lines also
include the corrections to the von-Schweidler law. The dashed horizontal lines indicate
the fit values for $f_q^\mr{sc}$ at the respective momentum transfers.}
\label{q=all.T=0.47}
\end{figure}

\begin{figure}
\begin{center}
\begin{minipage}[t]{100mm}
\epsfysize=90mm
\epsffile{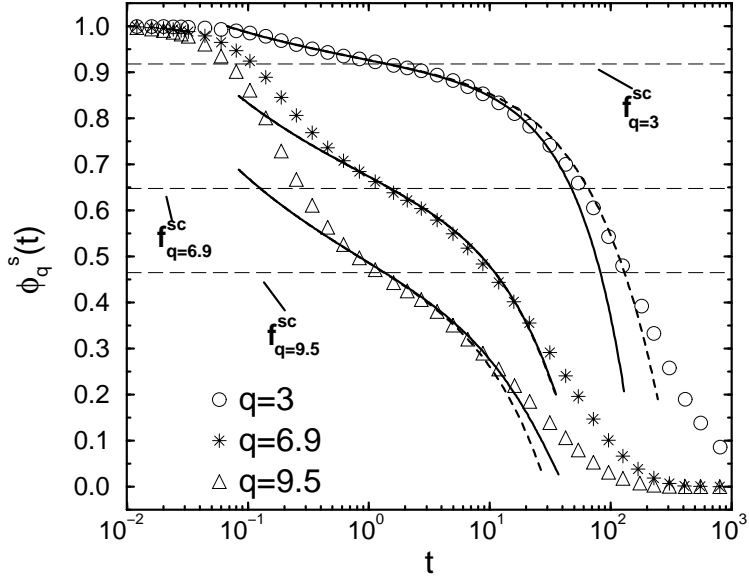}
\end{minipage}
\end{center}
\caption[]{Incoherent intermediate scattering function versus time at $T=0.52$. Three
$q$-values are shown: $q=3$ ($\approx$ size of the chain), $q=6.9$ [$\approx$ maximum of
$S(q)$], and $q=9.5$ [$\approx$ first minimum of $S(q)$]. The dashed lines are the 
MCT-fit results, using only the factorization theorem, whereas the solid lines also
include the corrections to the von-Schweidler law. The dashed horizontal lines indicate
the fit values for $f_q^\mr{sc}$ at the respective wave-vectors.}
\label{q=all.T=0.52}
\end{figure}

Figures~\ref{q=all.T=0.47} and \ref{q=all.T=0.52} 
show comparisons between the simulation data and 
Eq.~(\ref{eq2}) for $T=0.47$ and $T=0.52$, respectively.
For both temperatures the description of the data by 
the theory starts at about the same time, $t \approx 0.6$,
(except at $T=0.52$ and $q=3$, where the fits extends to
$t \leq 10^{-1}$).
This suggests that the scale $t_0$, which MCT introduces
as a matching time of Eq.~(\ref{eq2})
to the transient microscopics close to $T_\mr{c}$ ({i.e.}, 
$t_0\approx t_0(T_\mr{c}) = \mbox{constant}$), is 
only weakly -- if at all -- temperature dependent for our
polymer model. The same conclusion could have also been 
drawn from Fig.~\ref{q=all.T=0.47+0.52} because the 
scattering functions, calculated at different temperatures,
coincide for small times ({i.e.}, for $t < 0.3$).
Whereas the asymptotic result, 
$t_0\approx \mbox{constant}$, appears to hold quite generally
in experiments and simulations of nonpolymeric systems \cite{yip}, 
a pronounced temperature dependence 
was found in a Monte Carlo simulation of the bond-fluctuation 
model \cite{exmct}. Compared to the present study, this
seems to be a model-specific rather than a typical
feature of glass forming polymers, contrary to the 
conjecture of Ref.~\cite{exmct}.

For $t > 0.6$, the idealized theory describes the decay of
the correlators over about 1.5 decades in time at $T=0.52$ and 
over more than 2 decades at $T=0.47$. Therefore the 
$\beta$-window expands considerably in this narrow 
temperature interval. The extension of the window can also
be inferred from the shift of the time $\tau_\mr{co}$, where 
$f_q^\mr{sc}$ crosses the simulation data. For all $q$-values 
it increases from about $1.4$ at $T=0.52$ to about $10$ at 
$T=0.47$. The independence of $\tau_\mr{co}$ on $q$ is
an evidence of the factorization theorem which
implies $\phi_q^\mr{s} (\tau_\mr{co})=f_q^\mr{sc}$ for all
$q$, if $G(\tau_\mr{co})=0$. As $T \rightarrow T_\mr{c}^+$,
one expects $G(t)=0$ to occur close to the $\beta$-relaxation 
time so that $\tau_\mr{co} \propto t_\varepsilon$ \cite{ffgms}.
For $T>T_\mr{c}$, $t_\varepsilon$ marks the crossover of 
the critical to the von-Schweidler dynamics, where 
corrections are not dominant yet. This can be seen in
Figs.~\ref{q=all.T=0.47} and \ref{q=all.T=0.52}. The
idealized fits with and without corrections coincide on
the scale $\tau_\mr{co}$, but deviate at later times. 
Depending on $q$, corrections to the von-Schweidler law
extend the fits by about 0.25 decades at $T=0.52$ and 
by about 0.5 decades at $T=0.47$. For $q=3$, they are
negative, but positive for $q=9.5$. The change of sign
occurs approximately around $q=6.9$. This behavior 
qualitatively agrees with theoretical calculations for
hard spheres \cite{fgm} and recent simulations for linear
molecules \cite{kks}. 

Despite this agreement, we want to mention a problem that
we faced when applying the corrections. In the analysis 
we treated $\tilde{h}_q^\mr{s}$ and $\tilde{h}_q^{\mr{s}(2)}$ 
as temperature- and $q$-dependent fit parameters. However,
contrary to $\tilde{h}_q^\mr{s}$, the resulting temperature
dependence of $\tilde{h}_q^{\mr{s}(2)}$ turned out to be
rather irregular and sometimes even opposite to the theoretical 
expectation, especially at low temperatures. From Eq.~(\ref{eq2})
one expects the significance of the corrections to diminish
as $T_\mr{c}$ is approached, whereas Figs.~\ref{q=all.T=0.47}
and \ref{q=all.T=0.52} show that they become more important 
with decreasing temperature. This discrepancy could have
two reasons. First, $\tilde{h}_q^{\mr{s}(2)}$ is about
2 to 3 orders of magnitude smaller than $\tilde{h}_q^\mr{s}$
(which itself is of the order $10^{-2}$). Therefore it is
numerically difficult to extract reliable values from the 
fits, especially at large $q$-values, where $f_q^\mr{sc}$ is
small. Second, the idealized MCT is only supposed to work
in a temperature interval which is close, but not too close
to $T_\mr{c}$. In the immediate vicinity of $T_\mr{c}$,
\begin{figure}
\begin{center}
\begin{minipage}[t]{100mm}
\epsfysize=90mm
\epsffile{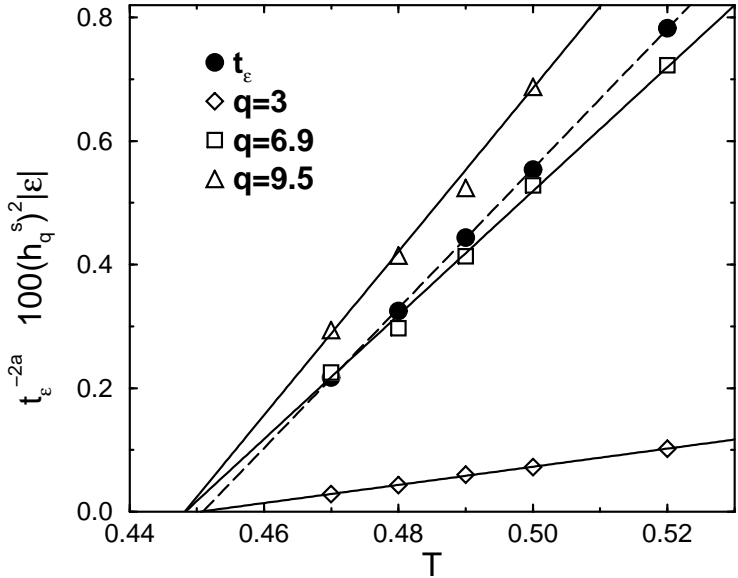}
\end{minipage}
\end{center}
\caption[]{Temperature dependence of the $\beta$-time scale, $t_\varepsilon$, and of the 
total prefactor of the $\beta$-scaling function, $\tilde{h}_q^\mr{s}=h_q^\mr{s}
|\varepsilon|^{1/2}$. Plots of $t^{-2a}_\varepsilon$ and of $(\tilde{h}_q^\mr{s})^2$
versus $T$ should extrapolate to zero at $T_\mr{c}$ [see Eq.~(\ref{eq3})]. The fit results 
(dashed line for $t_\varepsilon$ and solid lines for $\tilde{h}_q^\mr{s}$) can be 
combined to a common estimate: $T_\mr{c}=0.450 \pm 0.005$.}
\label{te+hq}
\end{figure}
ergodicity restoring hopping processes \cite{goetze1,fghl}
start to compete with the ergodicity breaking cage effect
that is treated by the idealized MCT. Qualitatively, the
main influence of the hopping processes is to accelerate
the decay of the correlators at late times. Since the
uncorrected idealized fits for small $q$ lie above the
simulation data, the addition of corrections to the 
von-Schweidler law could partly mimic the effect of 
hopping. This ambiguity can complicate the analysis.
Presumably, it would have been better to impose the
theoretical temperature dependence on $h_q^{\mr{s}(2)}$,
and to adjust only the variation with $q$. 

Despite this proviso, the idealized analysis should yield
reliable results on the $\beta$-time scale around 
$\tau_\mr{co}$, where neither corrections, nor hopping 
processes are dominant. From the fits one can extract
the $q$-dependences of $f_q^\mr{sc}$ and $h_q^\mr{s}$, and
the critical temperature. These results are shown in 
\begin{figure}
\begin{center}
\begin{minipage}[t]{100mm}
\epsfysize=90mm
\epsffile{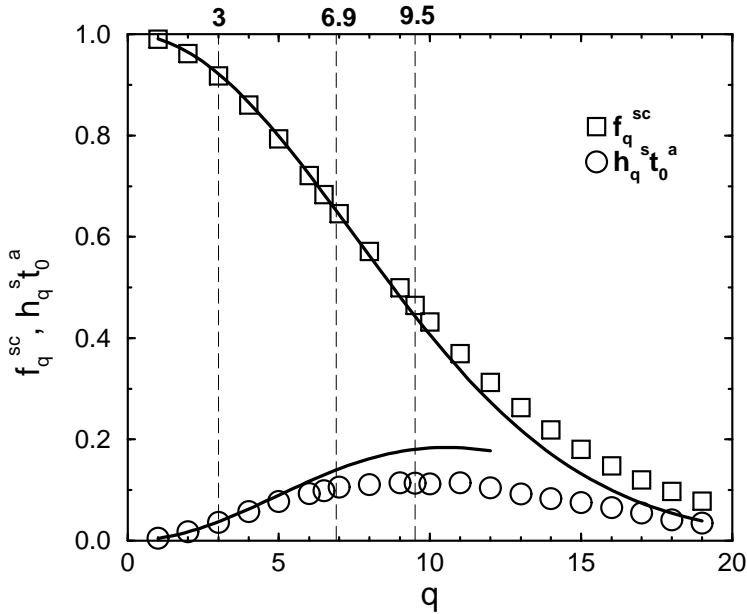}
\end{minipage}
\end{center}
\caption[]{$q$-dependence of the nonergodicity parameter $f_q^\mr{sc}$ and of the
critical amplitude $h_q^\mr{s}$ (in fact, $h_q^\mr{s}t_0^a$ is shown. The microscopic
matching time $t_0$ cannot be separated in the fit). The solid
curves are the Gaussian approximations of Eq.~(\ref{eq5}). The fit interval for 
$f_q^\mr{sc}$ was $1 \leq q \leq 8$. The result of this fit was used to estimate
$h_\mr{msd}$ from the behavior of $h_q^\mr{s}$ at small $q$.}
\label{fc+hq}
\end{figure}
Figs.~\ref{te+hq} and \ref{fc+hq}.  Figure~\ref{te+hq} plots 
$t_\varepsilon^{-2a}$ and $(\tilde{h}_q^\mr{s})^2$
versus temperature. The theoretically predicted linear behavior 
[see Eq.~(\ref{eq3})] is nicely confirmed by the data. A
linear regression yields $T_\mr{c}\simeq 0.451$ for $t_\varepsilon$.
Similar estimates are obtained for $\tilde{h}_q^\mr{s}$: 
$T_\mr{c}\simeq 0.451$ ($q=3$), $T_\mr{c}\simeq 0.448$ ($q=6.9$), 
and $T_\mr{c}\simeq 0.448$ ($q=9.5$). Combining these results
we find $T_\mr{c}=0.450 \pm 0.005$, which implies $0.04 \leq 
(T-T_\mr{c})/T_\mr{c} \leq 0.15$ for the reduced distance to the 
critical point in our model. Although $(T-T_\mr{c})/T_\mr{c}=0.04$ 
corresponds to the upper limit, where some of the asymptotic laws 
begin to be detectable in theoretical calculations for hard spheres 
\cite{fgm,ffgms}, similar reduced distances are not unusual in 
practical applications of MCT (see Refs.~\cite{cldhs,ka1}, for 
instance).

Figure~\ref{fc+hq} shows the $q$-dependences of $f_q^\mr{sc}$ 
and $h_q^\mr{s}$, together with the Gaussian approximations,
\begin{equation}
f_q^\mr{sc}\approx\exp[-q^2r_\mr{sc}^2] \quad \mbox{and} \quad
h_q^\mr{s}\approx h_\mr{msd}q^2\exp[-q^2r_\mr{sc}^2]\;,
\label{eq5}
\end{equation}
which are supposed to hold for small $q$ \cite{fgm}. As in
other experimental \cite{pbfksf} and theoretical studies \cite{kks,ka2},
the nonergodicity parameter of our model monotonously decreases with
increasing $q$, whereas the critical amplitude passes through
a maximum between the maximum and the first minimum of 
$S(q)$. The Gaussian approximations provide reasonable
descriptions for the initial behavior of both $f_q^\mr{sc}$ 
and $h_q^\mr{s}$. The description extends farther for 
$f_q^\mr{sc}$ ($q \leq 10$) than for $h_q^\mr{s}$ ($q \leq
4$), which is qualitatively comparable to the theoretical
results for hard spheres \cite{fgm,fhl}. From the fits one
obtains: $r_\mr{sc}=0.095\pm 0.005$ and $h_\mr{msd}t_0^a=0.0045
\pm 0.0010$. Both parameters enter the first two terms of 
the short-time expansion for the $\alpha$-process of the 
(monomer) mean-square displacement, which corresponds to
the von-Schweidler law in reciprocal space \cite{fgm}.
The mean-square displacement will be briefly discussed in 
the next section and more extensively together with related 
quantities in Ref.~\cite{bbpb_msd+rouse}.
\subsection{$\alpha$-relaxation regime}
\label{alpha}
The final structural decay from the nonergodicity parameter to zero
is called $\alpha$-relaxation regime. For this regime the idealized 
MCT makes the following predictions \cite{fgm,ffgms,fhl}. First, the
$\alpha$-process asymptotically obeys a time-temperature superposition 
principle, 
\begin{figure}
\begin{center}
\begin{minipage}[t]{100mm}
\epsfysize=90mm
\epsffile{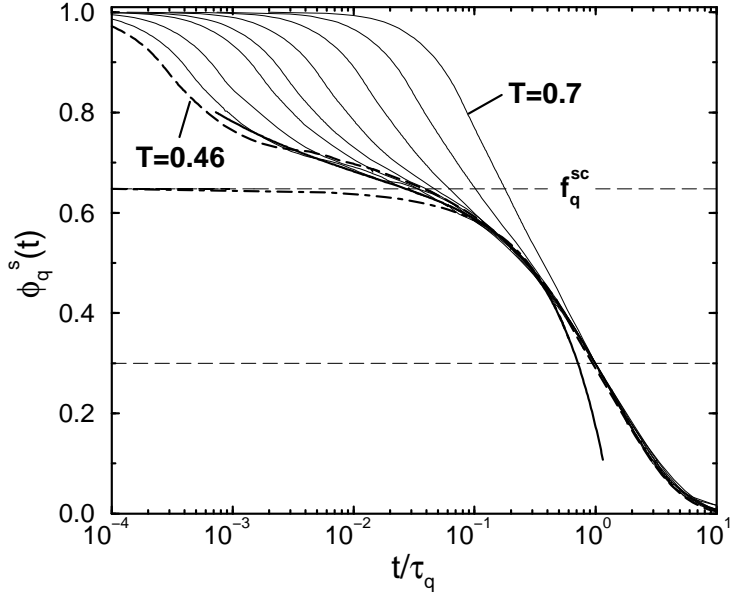}
\end{minipage}
\end{center}
\caption[]{$\alpha$-scaling plot of $\phi_q^\mr{s}(t)$ versus $t/\tau_q$
for $q=6.9$. Nine different temperatures are shown: $T=0.46,0.47,0.48,0.49,0.5,
0.52,0.55,0.6,0.7$ (from left to right). The scaling time $\tau_q$ was defined
by $\phi_q^\mr{s}(\tau_q)=0.3$ (lower horizontal dashed line). Note that the 
scaling extends to shorter
rescaled times with decreasing temperature if $T\geq 0.48$. For $T=0.46$
(dashed line) this trend is violated (see text for further discussion).}
\label{askale}
\end{figure}
\begin{equation}
\phi_q^\mr{s}(t)=\tilde{\phi}_q^\mr{s}(t/\tau) \; ,
\label{tts}
\end{equation}
{i.e.}, all correlators, measured for one $q$-value at different 
temperatures, should collapse onto a master curve if time is rescaled
by $\tau$ [see Eq.~(\ref{eq4})]. Second, the short-time expansion of
Eq.~(\ref{tts}) coincides with the long-time behavior of the 
$\beta$-process, {i.e.}, with the von-Schweidler law, $\phi_q^\mr{s}(t)=
f_q^\mr{sc}-h_q^\mr{s}B(t/\tau)^b$ ($B=0.476$ for our model). Third, the
leading-order corrections to Eq.~(\ref{tts}) extend the description 
beyond $f_q^\mr{sc}$ at short rescaled times. They are given by $\delta
\tilde{\phi}_q^\mr{s}(t/\tau)=h_q^\mr{s}(B_1/B)|\varepsilon|(\tau/t)^b$
(with $B_1=0.185$), and are identical to the corrections of the $\beta$-correlator
for $t\gg t_\varepsilon$. Forth, in the limit of large $q$ the $\alpha$ master
curve is given by a Kohlrausch function \cite{mf_ali93}
\begin{equation}
\phi_q^\mr{s}(t)=f_q^\mr{K}\exp \left [-\left(\frac{t}{\tau_q^\mr{K}}
\right)^{\beta_\mr{K}}\right] 
\label{kww}
\end{equation}
with $f_q^\mr{K}=f_q^\mr{sc}$, $\beta_\mr{K}=b$, and $\tau_q^\mr{K}(T)=
\hat{\tau}_q^\mr{K}\tau(T)$, where $\hat{\tau}_q^\mr{K} \propto q^{-1/b}$. On the other 
hand, it is found that Eq.~(\ref{kww}) represents a good approximation for the 
$\alpha$ master function even if $q$ is small \cite{fhl}. 

Figure~\ref{askale} shows a test of these predictions for $q=6.9$. A
scaling time $\tau_q$ was defined by $\phi_q^\mr{s}(\tau_q)=0.3$. This
is legitimate because any time belonging to the $\alpha$-regime is
expected to exhibit asymptotically the same temperature dependence,
{i.e.}, $\tau_q \sim \tau$, due to time-temperature superposition.
The figure shows that the time-temperature superposition property is borne out by the 
simulation data for $T \leq 0.7$. Therefore, it already starts a 
higher temperatures compared to the $\beta$-scaling. Theoretically,
such a difference can be rationalized by the fact that the corrections
to the $\beta$ master function are of order $\varepsilon^{1/2}$, and
thus larger than those of Eq.~(\ref{tts}), which are of order
$\varepsilon$ only. At $T=0.7$ the $\alpha$-scaling is realized for about the last 
$40\%$ of the decay. With decreasing temperature it extends to smaller rescaled
times in such a way that it progressively adjusts to the von-Schweidler law,
and deviations at shorter times are quantitatively described by the 
$\beta$-correlator. 

This qualitative trend persists as long
as $T \geq 0.47$. However, for $T=0.46$ the rescaled correlator moves away
from the von-Schweidler asymptote instead of approaching it further. This is
in contrast to the theoretical expectation. Although we could not 
propagate the chain over several times the radius of gyration, we believe
that the melt is equilibrated over the distances probed by $q=6.9$ so 
that the deviation is presumably not a residual nonequilibrium effect. 
A possible explanation is that the melt at $T=0.46$ ($(T-T_\mr{c})/T_\mr{c}=
0.0\bar{2}$) is already so close to $T_\mr{c}$ that the influence of ergodicity restoring
processes starts to be felt. These processes change the shape of the correlator 
in the von-Schweidler regime and prevent the strong increase of
the $\alpha$-relaxation time according to Eq.~(\ref{eq4}) from continuing
\cite{fghl}. It is therefore possible that a correlator falls above the 
idealized $\alpha$ master curve when trying to rescale it by a relaxation time 
which is too small. However, a comparable overshoot was not observed in a recent 
MD-simulation of diatomic molecules \cite{kks}, although the temperature dependence of 
the relaxation times deviated markedly from the idealized prediction at the lowest 
temperatures studied. Nevertheless, the low temperature curves start crowding in the 
plateau regime and no longer expand along the von-Schweidler asymptote for shorter
rescaled times. Therefore, it is conceivable that an overshoot could also occur
in this model at still lower temperatures.

\begin{figure}
\begin{center}
\begin{minipage}[t]{100mm}
\epsfysize=90mm
\epsffile{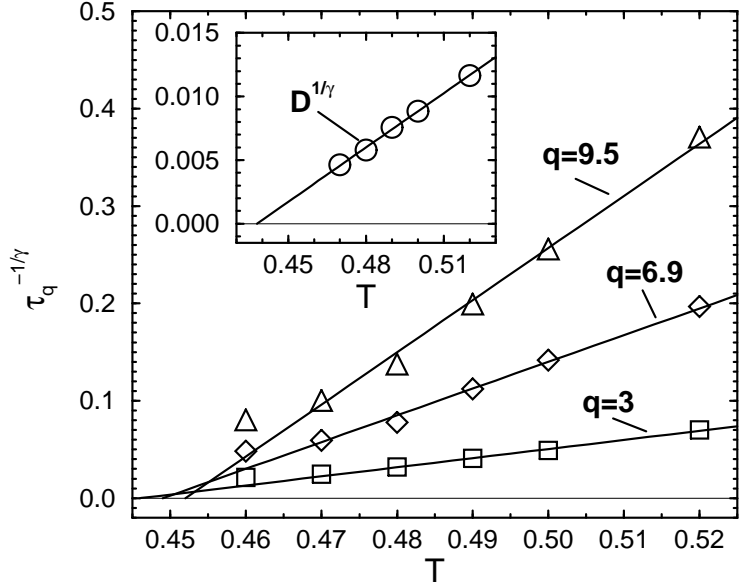}
\end{minipage}
\end{center}
\caption[]{Plot of $\tau_q^{-1/\gamma}$ versus temperature for $q=3$ ($\approx$
size of a chain), $q=6.9$ [$\approx$ maximum of $S(q)$], and $q=9.5$ [$\approx$
first minimum of $S(q)$]. The exponent $\gamma$ was taken from the
$\beta$-analysis ($\gamma=2.09$). The straight lines are fit results for the
interval $0.47 \leq T \leq 0.52$. The intersections of the fits with the
zero-line are estimates for $T_\mr{c}$: $T_\mr{c}=0.446$ ($q=3$),
$T_\mr{c}=0.449$ ($q=6.9$), $T_\mr{c}=0.452$ ($q=3$). The inset shows the
same plot for the diffusion constant. Here the fit result for the critical
temperature is significantly smaller, {i.e.}, $T_\mr{c}=0.438$.}
\label{t_gamma2}
\end{figure}

Figure~\ref{t_gamma2} shows that the relaxation time at $T=0.46$ is indeed 
smaller than expected from the mode-coupling fit.
The figure depicts a plot of $\tau_q^{-1/\gamma}$ and 
$D^{1/\gamma}$ ($D$: diffusion coefficient of a chain) versus $T$, using the 
best estimate for $\gamma$ from the $\beta$-analysis ($\gamma=2.09$). In the 
temperature interval, in which the idealized analysis was possible ($0.47 \leq T 
\leq 0.52$; at $T=0.46$ the $\beta$-analysis with the parameters quoted above
did not yield acceptable results), 
the simulation results lie on straight lines, as predicted by MCT. However, 
for $T=0.46$, the relaxation times are too small (no reliable estimate of 
the diffusion coefficient could be obtained after $5 \times 10^7$ MD-steps). 
The same observations were also made in Ref.~\cite{kks} for comparable reduced distances 
to the critical point.
Therefore the interval $0.47 \leq T \leq 0.52$ was used to obtain further estimates 
for $T_\mr{c}$. For $\tau_q$ this yields $T_\mr{c}\approx 0.446$ for $q=3$, 
$T_\mr{c}\approx 0.449$ for $q=6.9$, $T_\mr{c}\approx 0.452$ for $q=9.5$, and 
for $D$, $T_\mr{c}\approx 0.438$. Whereas the variation of $T_\mr{c}$ for 
$\tau_q$, albeit decreasing systematically with decreasing $q$, is compatible, 
within the error bars, with $T_\mr{c}=0.450$ 
from the $\beta$-analysis, the result from $D$ is too small. Barring the 
problem that the diffusion coefficient is hard to determine accurately at low
temperatures ({i.e.}, for $T \leq 0.47$), the high temperature behavior 
({i.e.}, $0.48 \leq T \leq 0.52$) suggests that the disparity in the 
$T_\mr{c}$-estimates is significant. Such a finding is not uncommon. Similar
observations were made in MD-simulations of a water model \cite{sfct} and of a
binary Lennard-Jones mixture \cite{ka1}. Whereas the difference in $T_\mr{c}$ in 
the water simulation was small and could be attributed to numerical uncertainties,
a much larger reduction was found for the binary mixture.
On the other hand, an unconstrained fit yielded a critical temperature which 
was compatible with the estimates from coherent and incoherent scattering, but
$\gamma$ was in turn significantly smaller than that from the $\beta$-analysis.
The same result was also obtained in MD-simulations of diatomic molecules
\cite{kks,kks1} and for our model \cite{bpbb_pressure}.

In experiments and simulations it is generally found that
the Kohlrausch function represents a very good fit formula
for the major part of the $\alpha$-relaxation. Only at 
early times deviations (can) become visible. Mode-coupling
theory interpretes these deviations as a signature of the
von-Schweidler law, since it concatenates the $\beta$- with 
the $\alpha$-process, and $\beta_q\neq b$ in general
\begin{figure}
\begin{center}
\begin{minipage}[t]{100mm}
\epsfysize=90mm
\epsffile{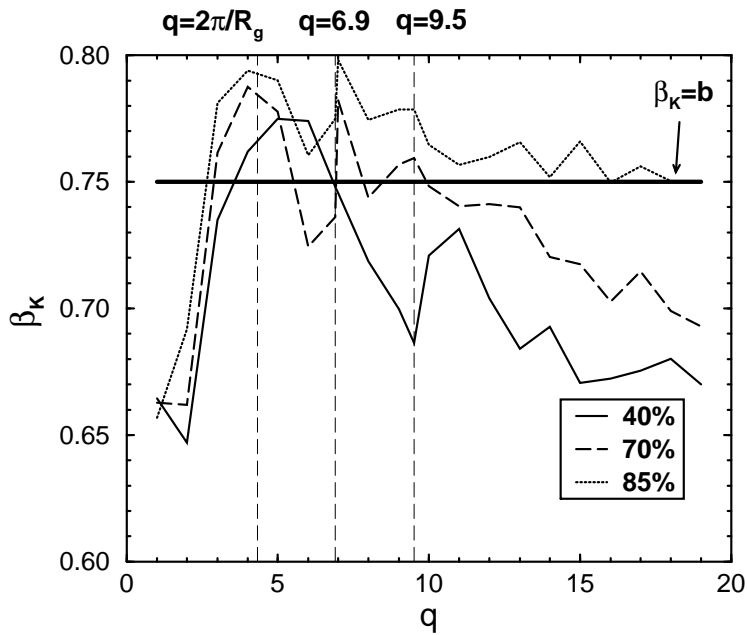}
\end{minipage}
\end{center}
\caption[]{Kohlrausch stretching exponent $\beta_\mr{K}$ [see Eq.~(\ref{kww})] as
a function of $q$ for three different choices of the fit interval 
$0.02 \leq \phi_q^\mr{s}(t)/f_q^\mr{sc} \leq X$ for $X=0.4, 0.7$ and $0.85$.
The thick horizontal line indicates
the von-Schweidler exponent $b=0.75$. In addition, three different $q$-values
are highlighted by vertical dashed lines: $q=2\pi/R_\mr{g}\approx 4.35$ ($R_\mr{g}$ =
radius of gyration), $q=6.9$ [$\approx$ maximum of $S(q)$], and $q=9.5$ [$\approx$
first minimum of $S(q)$].}
\label{beta}
\end{figure}
\cite{fhl}. Therefore, when applying the Kohlrausch function, the problem 
arises that the late $\beta$-process interferes with the fit \cite{fghl1}, 
and additionally, that the 
fit parameters sensitively depend on the size of the fit interval (for
a discussion of this problem see Refs.~\cite{lw,cdfghllt},
for instance).
To lessen this problem two procedures are often used. 
On the one hand, one can fit the late-time decay of the
$\alpha$ master function by taking into account that
the short-time bound of the fit interval should not 
overlap too much with the von-Schweidler regime. This procedure has been used in
Refs.~\cite{kks,ka1}, for instance.
On the other hand, one can work with so-called 
$\alpha$-$\beta$-fits, in which Eq.~(\ref{eq2}) (without
corrections) and Eq.~({\ref{tts}) are superimposed. 
Except for hard-sphere colloidal systems, where it is 
possible to calculate the master function numerically \cite{fhl,vmu},
Eq.~(\ref{tts}) is usually replaced by Eq.~(\ref{kww}). 
The superposition requires $f_q^\mr{K}=f_q^\mr{sc}$,
and the von-Schweidler law to be subtracted from the
$\beta$-correlator because it is already approximated by
the Kohlrausch function \cite{fcdgllt,bjncs_95} (for an
alternative procedure see \cite{cldhs}).

A variant of the latter approach was applied in this 
simulation. We only worked with the Kohlrausch function, 
but fixed the amplitude by $f_q^\mr{K}= f_q^\mr{sc}$. Then 
we tried to optimize the stretching exponent $\beta_{\rm K}$ (the
time scale is already given by $\phi_q^\mr{s}(\tau_q^\mr{K})=f_q^\mr{sc}\mr{e}^{-1}$). 
The results are still very sensitive to the choice of the fit 
interval. An example is given in Fig.~\ref{beta}. It shows 
the $q$-dependence of the stretching exponent 
$\beta_{\rm K}$. The percentages specify the portion of the 
decay from $f_q^\mr{sc}$ to (about) 0, 
which is included in the fit 
({i.e.}, ``$40\%$'' means the range $0.02 \leq 
\phi_q^\mr{s}(t)/f_q^\mr{sc} \leq 0.4$). 
\begin{figure}
\begin{center}
\begin{minipage}[t]{100mm}
\epsfysize=90mm
\epsffile{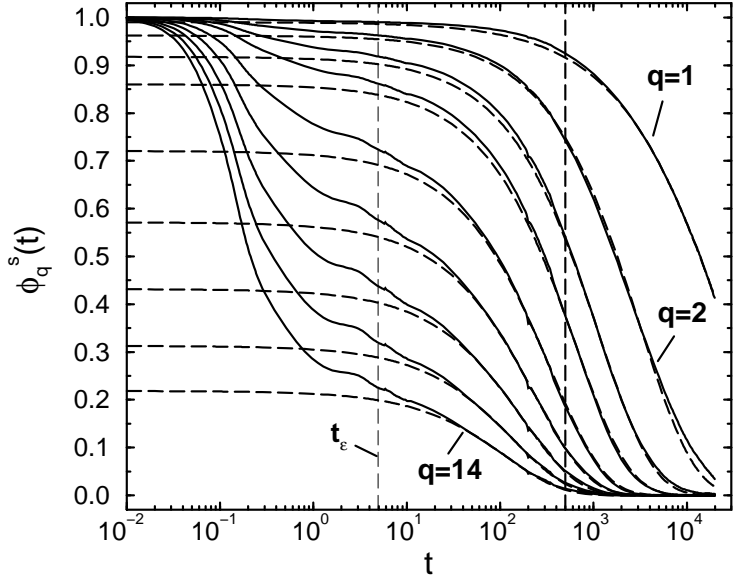}
\end{minipage}
\end{center}
\caption[]{$\phi_q^\mr{s}(t)$ and Kohlrausch functions at $T=0.48$. Solid lines are the 
simulation data for nine different $q$-values: $q=1,2,3,4,6,8,10,12,14$ 
(from right to left). For $q\geq 2$, the Kohlrausch parameters are given by:
$f_q^\mr{K}=f_q^\mr{sc}$, $\beta_\mr{K}=b=0.75$, and $\tau_q^\mr{K}$ from $\phi_q^\mr{s}
(\tau_q^\mr{K})=f_q^\mr{sc}\mr{e}^{-1}$. These choices provide a good description of 
the $\alpha$-decay, as long as $q > 2$. For $q\leq 2$, deviations gradually develop. The
stretching becomes more pronounced, leading to the following parameters at $q=1$:
$f_q^\mr{K}=f_q^\mr{sc}$, $\beta_\mr{K}=0.656$, and $\tau_q^\mr{K}=24595$. In addition,
two vertical lines are shown. The thin dashed line is $t_\epsilon(T=0.48)=4.933$,
whereas the thick dashed line indicates the lower time value ($t=500$), above which the
monomer mean-square displacement exhibits a $t^{0.65}$-behavior (see Fig.~\ref{qall+gauss.T=0.48}).}
\label{co+kww.qall.T=0.48}
\end{figure}
At large $q$ the nonergodicity parameter is so small that
statistical noise of the data considerably influences
the results. The $\beta_\mr{K}$-values for the various fit 
intervals then strongly splay out, leading to a difference 
of about $15\%$ at $q=19$. With decreasing $q$ the curves
approach one another and fluctuate around $\beta_\mr{K}
\approx b=0.75$, except for the smallest $q$-values, 
$q=1,2$, where $\beta_\mr{K}$ lies between 0.65 and 0.7.

This behavior suggests that it should be possible to find
a fit interval for $q > 2$, where the Kohlrausch function
with $f_q^\mr{K}=f_q^\mr{sc}$ and $\beta_\mr{K}=b$ 
provides a good description of the $\alpha$-process. 
Figure~\ref{co+kww.qall.T=0.48} shows a comparison between 
$\phi_q^\mr{s}(t)$ and such Kohlrausch functions for a selection 
of $q$-values, ranging from $q=1$ to $q=14$, at $T=0.48$. One can 
clearly see that the expectation $\beta_\mr{K}=b$ is well borne out
if $q \geq 3$, but that smaller stretching exponents are required 
for $q < 3$. 

In addition, the data exhibit a small bump at about $t=3$ for all wave 
vectors. This bump gradually develops at low temperatures ($T < 0.52$), 
but is hard to see in the presentation of Figs.~\ref{q=all.T=0.47+0.52} 
or \ref{q=all.T=0.47}. A similar feature was also observed
in other simulations of fragile glass formers, for instance, for 
orthoterphenyl \cite{lw} or for a binary Lennard-Jones mixture \cite{ka2},
and is much more pronounced for strong glasses, like silica
\cite{jkba}. Usually, the bump is attributed to a finite size
effect. Due to nonlinear coupling with other wave vectors a sound
wave, which propagates through the simulation box, leaves it on one
side, and immediately reenters the box on the other side because of
periodic boundary conditions, could generate a disturbance at a
time $t \approx L/c$, where $L$ is the linear dimension of the box
and $c$ is the sound velocity. Using $L=10.5$, and estimating the
sound velocity as $c \approx 7$,\footnote{The sound velocity is 
estimated from $c=(c_p/c_V\rho\kappa_T)^{1/2}$ by taking into account 
$S(q\rightarrow 0)=T\rho \kappa_T$ and the thermodynamic relation between 
the specific heats, $c_p$ and $c_V$, and the thermal expansion coefficient.}
we find $t \approx 1.5$, which is close to the position of the bump.

Two comments with respect to the Kohlrausch fits have to be made.
First, the equality $\beta_\mr{K}=b$ starts to work at
$q$-values which are about a factor of 2 smaller than
the maximum position of $S(q)$. Such a behavior has neither
been observed in experiments \cite{fcdgllt,bjncs_95} 
nor in simulations \cite{ka1} of nonpolymeric glass formers. 
Theoretically, one expects
$\beta_\mr{K}=b$ to become valid if $q\rightarrow \infty$, 
and numerical calculations for hard spheres indicate that 
this limit is approximately realized only for $q$-values 
which are 5--6 times larger than the maximum position of 
$S(q)$ \cite{fhl,mf_ali93}. Therefore the present finding seems
to be a special property of the model under consideration,
although neutron-scattering experiments for polybutadiene
\cite{zrff,zorn97} and other polyolefines \cite{col95,col93}
also suggest $\beta_\mr{K}={\rm const}$ (and perhaps $\approx b$ for
polybutadiene \cite{zrff}) for $q$-values that
lie around the maximum of $S(q)$ (typically, $2\pi/R_\mr{g} \ll
0.2 \leq q \leq 5$ ${\rm \AA}^{-1}$).

The second comment concerns the drop of $\beta_\mr{K}$ for
small $q$. Usually, one expects $\beta_\mr{K}$ to
increase monotonously to 1 as $q\rightarrow 0$ because
the system becomes freely diffusive on this length scale. 
This behavior is well borne out in theoretical calculations
for hard spheres \cite{fhl} and simulations of simple liquids
\cite{ka2}.
We believe that the difference observed here is related to 
the polymeric character of our model because $q\leq 2$
probes the length scale of a chain. For these $q$-values
the Gaussian approximation,
\begin{equation}
\phi_q^\mr{s}(t)=\exp\left [-\frac{1}{6}q^2 g_0(t)\right] \;,
\label{phigau}
\end{equation}
represents a (very) good description of the data
(see Fig.~\ref{qall+gauss.T=0.48}) so that one can 
\begin{figure}
\vspace*{-15mm}
\begin{center}
\begin{minipage}[t]{100mm}
\epsfysize=90mm
\epsffile{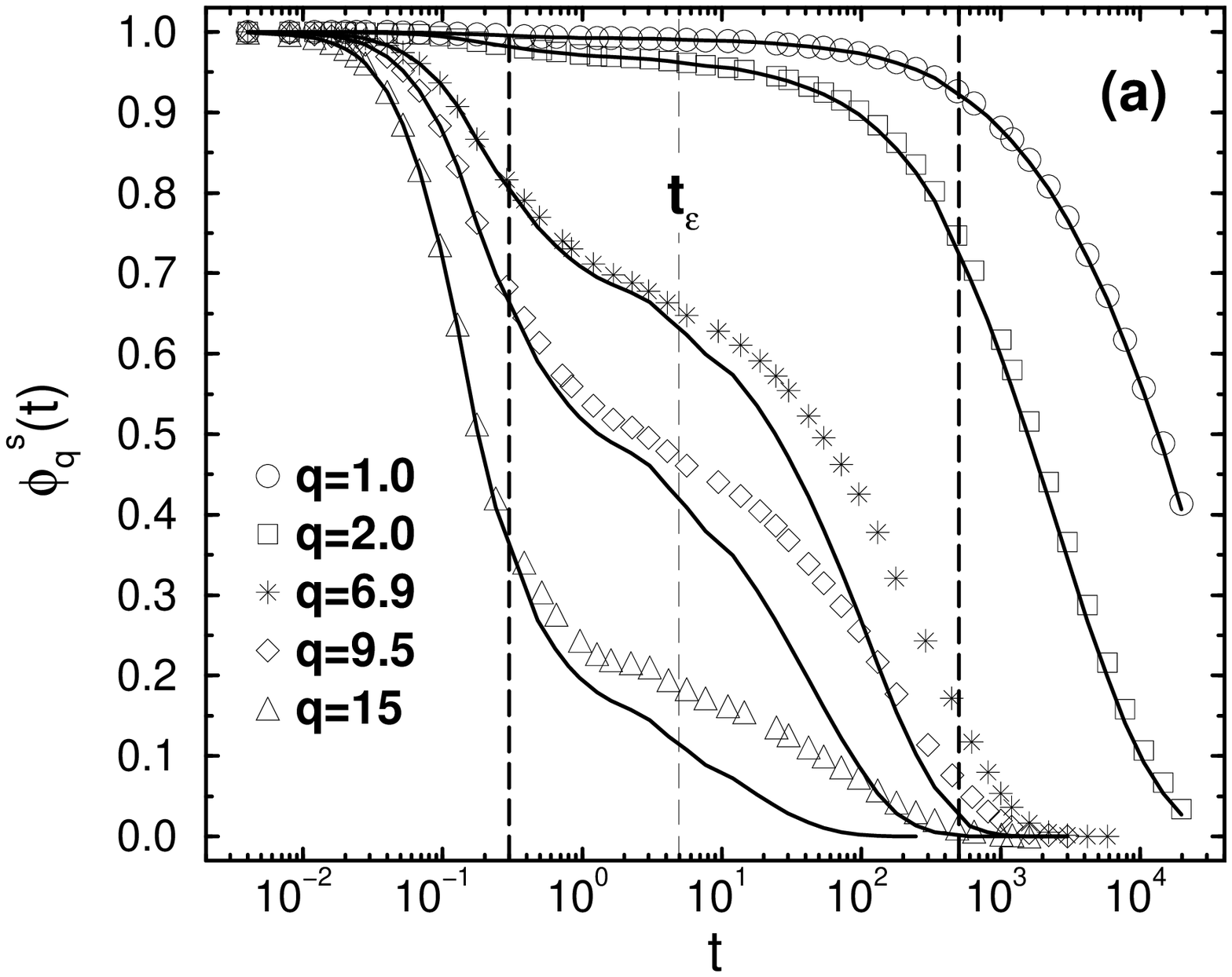}
\vspace*{-10mm}
\epsfysize=90mm
\epsffile{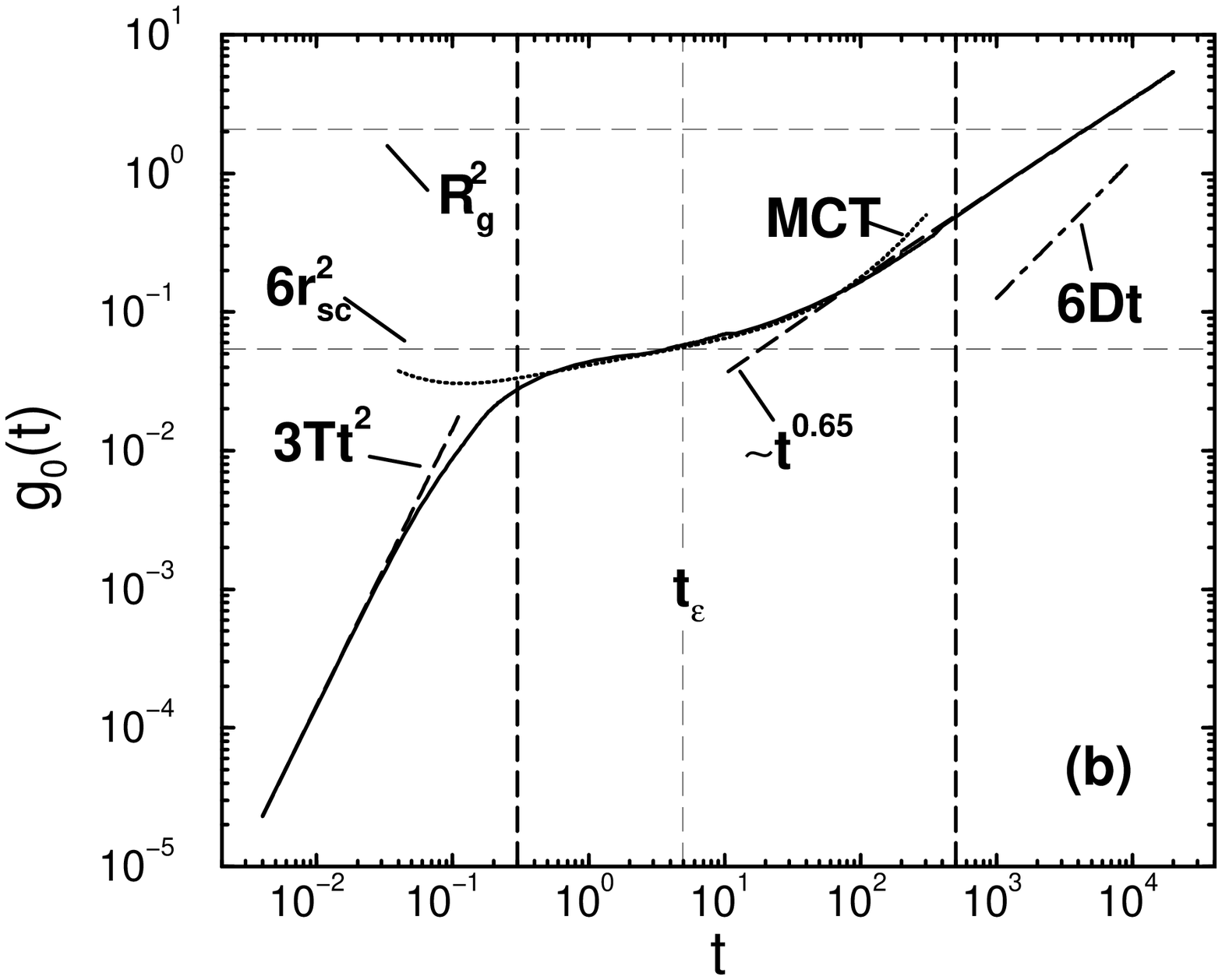}
\end{minipage}
\end{center}
\caption[]{Panel (a) shows a comparison between $\phi_q^\mr{s}(t)$ (symbols) and the 
Gaussian approximation Eq.~(\ref{phigau}) (solid lines) for five different $q$-values at 
$T=0.48$. The $q$-values smaller than 3 probe the size of a chain (see Fig.~\ref{q_joerg}). 
For $q=1$ ($< 2\pi/R_\mr{e}$)
the agreement is quantitative. Panel (b) shows the mean-square displacement, $g_0(t)$,
of all monomers (solid line) at $T=0.48$. The initial slope is $g_0(t)=3Tt^2=1.44 t^2$ (dashed
line). The dotted line is a fit by Eq.~(\ref{msdmct}) with $6r_\mr{sc}^2=0.054$ (dashed horizontal
line). In addition, a power law, $g_0(t) \sim t^{x_0}$ with $x_0=0.65$, is indicated by another 
dashed line. To determine $x_0$, only the simulation data larger than $R_\mr{g}^2=2.09$ were used.
Nevertheless, the power law extends by about one decade to smaller times. The dash-dotted line
shows the expected long time behavior $g_0(t)=6Dt$. The diffusion coefficient was determined
from the mean-square displacement of a chain, which reaches the diffusive limit earlier than
$g_0(t)$ \cite{bbpb_msd+rouse}. In both panels three vertical dashed lines are shown. From left 
to right, the
first ($t=0.3$) indicates the approximate time, when the Gaussian approximation stops working,
if $q \geq 5$, the second is the $\beta$-time scale $t_\epsilon(T=0.48)=4.933$,
and the last marks the onset of the $t^{0.65}$ power law.} 
\label{qall+gauss.T=0.48}
\end{figure}
interprete the decay of $\phi_q^\mr{s}(t)$ in terms of 
the monomer mean-square displacement\footnote{All monomers are 
included in $g_0(t)$, as it was the case in the calculation of 
$\phi_q^\mr{s}(t)$. For a further discussion of this and other mean-square
displacements see Ref.~\cite{bbpb_msd+rouse}.} $g_0(t)$. At early
times ($t\leq 0.03$) the mean-square displacement is purely
ballistic, {i.e.}, $g_0(t)=3Tt^2=1.44t^2$ ($T=0.48$), before it
crosses over to a plateau regime at about $t\approx 0.3$. At this time,
Eq.~(\ref{phigau}) also ceases to be a good approximation for $\phi_q^\mr{s}(t)$, 
if $q$ is too large ({i.e.}, for $q \geq 5$). The 
leveling-off of $g_0$ reflects the temporary localization
of the monomers in their cages. It is the counterpart of
the $\beta$-relaxation in real space, and can be described
by \cite{fgm}
\begin{equation}
g_0(t)\simeq
6r_\mr{sc}^2-6h_\mr{msd}\left [\frac{t_0}{t_\varepsilon}\right]^{a} g(t/t_\varepsilon)
-6h_\mr{msd}C_a\left [ \frac{t_0}{t}\right]^{2a} 
-6h_\mr{msd}B^2C_b\left [ \frac{t_0}{t_\varepsilon}\right]^{2a}
\left [ \frac{t}{t_\varepsilon}\right]^{2b} \; ,
\label{msdmct}
\end{equation}
where $C_a$ and $C_b$ are constants, which result from the corrections to the
critical decay and to the von-Schweidler law in the limit $q\rightarrow 0$, 
respectively. For hard spheres, one expects $C_a <0$ and $C_b < 0$ \cite{fgm,ffgms}.
Figure~\ref{qall+gauss.T=0.48}b includes a fit of Eq.~(\ref{msdmct}) to $g_0(t)$. Please 
note that $r_\mr{sc}^2=0.009$ ($r_\mr{sc}\simeq 0.095$ corresponds to the Lindemann 
criterion of melting), $h_\mr{msd}t_0^a=0.0045$, $t_\varepsilon=4.933$, $B=0.476$, and 
$g(\hat{t})$ are taken from the $\beta$-analysis. The only adjustable parameters are 
$C_at_0^{a}$ and $C_bt_0^{a}$. The fit yields the reasonable values $C_at_0^{a}\approx 
- 0.3$ and $C_bt_0^{a}\approx - 0.25$. On the other hand, it is also possible to fit
$g_0(t)$ by the leading-order ansatz $g_0(t)=6r^2_\mr{sc}+At^b$, where $r^2_\mr{sc}$
and $A$ have to be adjusted \cite{bbbp98} (see also Ref.~\cite{ka1} for another
application to simple liquids). Although such a fit extends the description of 
Eq.~(\ref{msdmct}) at late times by approximately one decade, it requires $r_\mr{sc} 
\approx 0.087$ and $A \approx 3.68 \times 10^{-3}$. These values are considerably 
different from those of the $\beta$-analysis. Therefore, the first approach should be
preferred.

Equation~(\ref{msdmct}) describes the simulation data from $t \approx 0.5$ to about 
$t \approx 100$, {i.e.}, up to the initial relaxation of a monomer out of its
cage. During the $\beta$- and the early $\alpha$-relaxation a monomer hardly feels the 
bonds to its neighbors along the chain. The mean-square displacement is of the order
$10^{-2}-10^{-1}$ of the monomer diameter. Therefore the monomer behaves like a particle 
in a simple liquid. For larger displacements chain connectivity
starts to influence the monomer dynamics, and finally 
becomes dominant. Whereas the von-Schweidler behavior
crosses over to free diffusion in simple liquids \cite{ka1}, 
a further subdiffusive regime, $g_0(t) \sim t^{x_0}$ 
($x_0 \approx 0.65$), intervenes for our model if 
$t \geq 500$. Such subdiffusive displacements are
well known in polymer simulations of nonentangled chains
\cite{bipa97}, and usually interpreted as a crossover
between a $t^{1/2}$-behavior -- the prediction of the
Rouse model \cite{doi} for $1 \ll g_0(t) \ll R_\mr{e}^2$ --
and free diffusion. A comparison of Figs.~\ref{co+kww.qall.T=0.48}
and \ref{qall+gauss.T=0.48} shows that
the subdiffusive displacement occurs in the same time
regime, where the stretched exponential with $\beta_\mr{K}=0.656$
fits $\phi_q^\mr{s}(t)$ for $q=1$.
Therefore, we suggest that the drop of $\beta_\mr{K}$ 
is a polymer specific effect, related to chain connectivity,
which becomes visible if $q$ probes the length scales 
between the monomer diameter ($=1$ in our units) and the 
end-to-end distance $R_\mr{e}$. On the other hand, if $q \ll
2\pi/R_\mr{e}$, the monomer displacement has to become diffusive,
and so $\beta_\mr{K}$ should approach 1, as in simple
liquids. Therefore we expect $\beta_\mr{K}$ to exhibit
a minimum for our model at small $q$.

Finally, Fig.~\ref{tau_of_q.allT} shows the $q$-dependence of the 
$\alpha$-relaxation time $\tau_q$. Here, $\tau_q$ was defined 
by $\phi_q^\mr{s}(\tau_q)=f_q^\mr{sc}/2$ instead of by $\tau_q^\mr{K}$
because $\tau_q^\mr{K}$ is larger than the simulated times if $q < 2$.
However, for $q$-values, where both times are available, we 
checked that they qualitatively yield the same result. In addition, 
$\tau_q$ was divided by $t_\varepsilon^{1+a/b}$ to eliminate the critical
dependence on $T_\mr{c}$ [see Eq.~(\ref{eq4})]. Asymptotically, the 
resulting quantity, $\tilde{\tau}_q=\hat{\tau}_qt_0^{-a/b}$ 
[see Eq.~(\ref{kww})], should be temperature 
independent. The figure shows that the division does not completely
remove the temperature dependence. For $q=1$ the relaxation times at 
$T=0.47$ and $T=0.52$ are about a factor of 2 different, with 
$\tilde{\tau}_q(T=0.47) < \tilde{\tau}_q(T=0.52)$, whereas both 
$\tilde{\tau}_q$'s first approach, and then cross one another around the
first peak of the static structure factor ($q\simeq 6.9$), before splaying out again. At
$q=19$ there is again about a factor of 2 between $T=0.47$ and $T=0.52$, but
now $\tilde{\tau}_q(T=0.47) > \tilde{\tau}_q(T=0.52)$. However, the absolute 
values of $\tilde{\tau}_q$ are of the order $10^{-1}$ for $q=19$, and thus
1--2 orders of magnitude smaller than $t_\varepsilon^{1+a/b}$, while they
are comparable to $t_\varepsilon^{1+a/b}$ at $q=1$. Therefore the difference
at small $q$ seems to be more important. It could again indicate that the relaxation 
times on the largest length scales do not increase as rapidly as expected from 
idealized MCT, an observation, which we have already made for the diffusion coefficient 
(see Fig.~\ref{t_gamma2} and Ref.~\cite{bpbb_pressure} for further discussion of this
point). On the other hand, it is not clear how significant the 
found deviations are at all, since the dominant temperature dependence has been taken 
into account by $t_\varepsilon^{1+a/b}$, and an additional smooth temperature 
dependence is expected, if $T$ is not very close to $T_\mr{c}$ (see Fig.~7 of 
Ref.~\cite{ffgms}, for instance).
\begin{figure}
\begin{center}
\begin{minipage}[t]{100mm}
\epsfysize=90mm
\epsffile{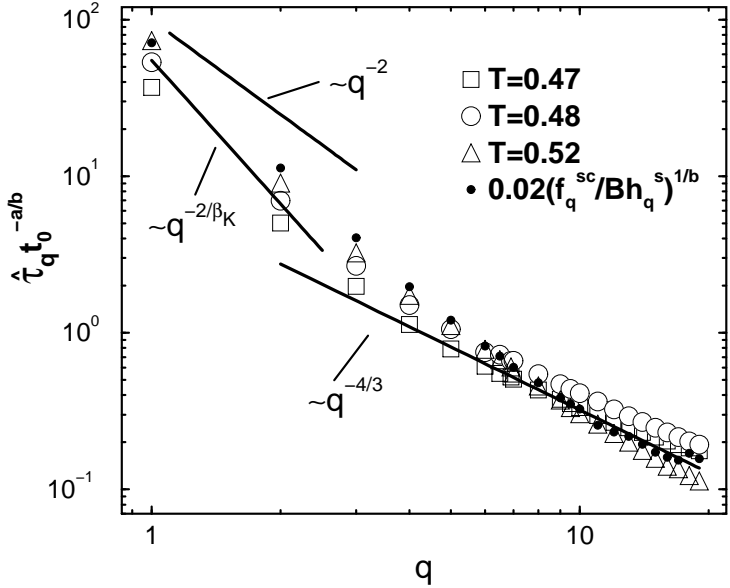}
\end{minipage}
\end{center}
\caption[]{$q$-dependence of the $\alpha$-relaxation time $\tau_q$, defined by $\phi_q^\mr{s}(
\tau_q)=f_q^\mr{sc}/2$. The figure shows $\hat{\tau}_qt_0^{-a/b} =\tau_q/t_\varepsilon^{1+a/b}$ 
[see Eq.~(\ref{eq4})] for three different temperatures from the interval, where the $\beta$-analysis
could be done. Thick solid lines indicate different asymptotic behavior. The initial decrease of
$\hat{\tau}_qt_0^{-a/b}$ is steeper than $q^{-2}$, the expected dependence for free diffusion, but
compatible with $q^{-2/\beta_\mr{K}}=q^{-3.05}$ ($\beta_\mr{K}=0.656$) [see Eq.~(\ref{tauqkww})]. 
For larger $q$, 
$\hat{\tau}_qt_0^{-a/b}$ crosses over to the mode-coupling result, Eq.~(\ref{tauq}). The filled 
points show $(f_q^\mr{sc}/Bh_q^\mr{s})^{1/b}$, whereas the solid line is the asymptotic limit,
$\sim q^{-1/b}$, for $q\rightarrow\infty$.}
\label{tau_of_q.allT}
\end{figure}

The $q$-dependence of the $\alpha$-relaxation time can be roughly divided into
two regions. For $q< 2$ Figs.~\ref{co+kww.qall.T=0.48} and \ref{qall+gauss.T=0.48}
showed that the Gaussian approximation, Eq.~(\ref{phigau}), and the Kohlrausch function
with $\beta_\mr{K}=0.656$ describe the $\alpha$-relaxation of $\phi_q^\mr{s}(t)$ well.
In the $\alpha$-regime one can therefore approximately equate Eqs.~(\ref{kww}) and 
(\ref{phigau}). Furthermore, if one assumes a power law for the $q$-dependence of $\tau_q$, 
one obtains
\begin{equation}
\tau_q \sim q^{-2/\beta_\mr{K}} = q^{-3.05} \quad \mbox{for} \quad q < 2 \;.
\label{tauqkww}
\end{equation}
Figure~\ref{tau_of_q.allT} shows that this estimate reasonably agrees with the 
initial behavior of $\tau_q$. Equation~(\ref{tauqkww}) has been suggested 
already some time ago in neutron-scattering experiments of glass-forming polymers
\cite{zrff,col95}, if $q^2g_0(t)$ is sufficiently small to warrant the Gaussian 
approximation (see also Ref.~\cite{zorn97} for a critical discussion of this issue). 
If $q \geq 5$, $\tilde{\tau}_q$ is indicative of another power-law behavior that is
compatible with the  mode-coupling prediction \cite{fhl,mf_ali93},
\begin{equation}
\hat{\tau}_q=\left[\frac{f_q^\mr{sc}}{Bh_q^\mr{s}}
\right]^{1/b}\stackrel{q\rightarrow\infty}{\longrightarrow}
q^{-1/b} \;, 
\label{tauq}
\end{equation}
which is expected to hold, if $\beta_\mr{K}=b$. A similar observation was made in 
neutron-scattering experiments of orthoterphenyl \cite{pbfksf}.
\section{Conclusions}
\label{sum}
This paper presents results of a molecular-dynamics
simulations for a simple model of a glassy, nonentangled polymer melt.
The discussion focuses on the monomer dynamics above
$T_\mr{c}$, as monitored by the incoherent intermediate
scattering function $\phi_q^\mr{s}(t)$ and the mean-square
displacement. These functions do not distinguish between 
bonded and nonbonded monomers so that it is {\em a priori} 
not clear how the polymeric character influences 
$\phi_q^\mr{s}(t)$ in the supercooled state. 
From the analysis the following picture emerges: 
\begin{enumerate}
\item 
At very early times the motion of a monomer is purely ballistic.
This corresponds to the phonon contribution in laboratory
glass formers. 
\item 
At later times, the ballistic motion
slows down because the monomer feels the confinement imposed 
by its nearest neighbors. The monomer becomes almost
localized, and the period of this localization extends
over about two decades in time in the narrow temperature interval
$0.47 \leq T \leq 0.52$. ``Localization'' in this context
means that the mean-square displacement remains close to
$r_\mr{sc}\approx 0.095$, {i.e.}, to a displacement
that is about $10\%$ of the monomer diameter. This is the
regime of the $\beta$-relaxation, which deals with the 
approach towards and leaving of the plateau value 
$r_\mr{sc}$ (see Fig.~\ref{qall+gauss.T=0.48}b).
In this sense, $r_\mr{sc}$ can be interpreted as a dynamic
measure of the size of the cage. Motions on this length
scale are so small that they are hardly affected by the polymeric
character, {i.e.}, by the nature of the glass former. 
This is presumably the reason, why mode-coupling theory,
a theory developed for simple liquids, can give a reasonable 
explanation for the dynamics of structurally much more
complicated systems in the $\beta$-regime. 
\item 
Concerning the comparison with the idealized MCT in
the $\beta$-regime, the simulation provides evidence for 
the space-time factorization theorem with temperature
independent $f_q^\mr{sc}$, $h_q^\mr{s}$, and $\lambda$,
as long as $0.47 \leq T\leq 0.52$. In this temperature 
region, the $\beta$-relaxation time exhibits the predicted 
power law, yielding a critical temperature of $T_\mr{c}
\simeq 0.45$, and the $q$-dependences of $f_q^\mr{sc}$ and
$h_q^\mr{s}$ are in qualitative agreement with calculations
for hard spheres. For $T<0.47$, deviations from the 
idealized behavior are observed, which can be interpreted
in terms of ergodicity restoring hopping 
processes. We have not attempted an extended MCT-analysis
because data for only one temperature below $T=0.47$
had been simulated. However, we can roughly estimate an upper
bound of the hopping parameter $\delta$ for our model.
Using $\delta t_0=(t_0/t_\varepsilon)^{1+2a}$ \cite{fghl},
$t_\varepsilon(T=0.47)=8.75$, and assuming $t_0 \approx 0.3$
(due to Fig.~\ref{qall+gauss.T=0.48}),
we obtain $\delta t_0 \approx 3 \times 10^{-3}$, which
would be larger than experimental values at similar 
distances from the critical point \cite{cldhs,cdfghllt}.
\item 
When the mean-square displacement becomes larger than $r_\mr{sc}$,
the monomer begins to leave its cage, and the temporarily
frozen structure ``melts'' (Lindemann criterion). The 
initial stage of this ``melting'', i.e., of the
$\alpha$-process, can be described by the von-Schweidler 
law. But for larger times, polymer specific properties start 
dominating the dynamics. For instance, the mean-square
displacement exhibits a subdiffusive behavior between the
von-Schweidler law and free diffusion due to chain
connectivity. The exponent of the corresponding power
law is typical of the short chains studied. It also determines
the Kohlrausch stretching exponent for small wave-vectors, where
the Gaussian approximation holds. Qualitatively, this difference between 
$\alpha$- and $\beta$-processes is expected by mode-coupling
theory because the space-time factorization of the 
$\beta$-regime, which implies the same dynamics on all
length scale, is no longer valid in the late 
$\alpha$-regime. However, the theory still predicts that
the diffusion coefficient and the $\alpha$-relaxation
time $\tau_q$ should exhibit the same temperature 
dependence. In this respect, we find deviations between 
theory and simulation. The diffusion coefficient does not
decrease as quickly as $\tau_q$ at the maximum of the
structure factor with decreasing temperature. Somehow the
melt stays more mobile on large length scales than on
local ones. On the other hand, we find evidence for the time-temperature
superposition principle for the $\alpha$-process, if $T\geq 0.47$.
\item
The wave-vector dependence of the $\alpha$-relaxation time consists of
two regimes. For small $q$, where the Gaussian approximation holds, we find 
a $q^{-2/\beta_\mr{K}}$-behavior, as in some neutron-scattering experiments 
\cite{zrff,col95}, which crosses over to a $q^{-1/b}$-behavior for $q \geq 5$.
The latter power law is a MCT-prediction for $q$-values much larger than
the maximum of $S(q)$. Why this asymptotic behavior can already be observed
for rather small $q$ in the present model, is not clear.
\end{enumerate}
In summary, the dynamics of our model in the supercooled state can be 
understood as an interplay of mode-coupling and polymer specific 
effects. The ``caging'' of a monomer by its neighbors leads to a temporary 
trapping of the monomers and to a concomittant slowing down of the structural 
relaxation, as in simple liquids. This cage effect is dominant, as long as 
the monomer displacement is small (i.e., much smaller than the diameter of a 
monomer). The polymeric character of the model only determines the nonuniversal 
parameters of MCT. However, if a monomer gradually leaves its cage, 
chain connectivity becomes more and more influential. For the 
present model of short nonentangled chains, it leads to a subdiffusive 
Rouse-like displacement. The subdiffusive behavior interferes with the
late-$\beta$/early-$\alpha$ dynamics and limits the von-Schweidler regime to
a range, which is much smaller than in simple Lennard-Jones liquids, before
free diffusion sets in.
%
%
\section*{Acknowledgement}
We are indebted to Prof.\ K. Binder and Drs.\ W. Kob, A. Latz
and B. D\"unweg for many helpful discussions, and to Prof.\ K.
Binder and Dr.\ W. Kob for a critical reading of the manuscript.
We also like to thank Dr.\ M. Fuchs for valuable comments 
which influenced the final presentation of the data very 
much. In the course of this work, we have profited
from generous grants of simulation time by the computer
center at the university of Mainz and the HLRZ J\"ulich,
which are gratefully acknowledged, as well as financial support 
by the Deutsche Forschungsgemeinschaft under SFB262/D2.
%
%


\begin{thebibliography}{99}
\bibitem{ali_93}
Proceedings of the 2nd International Discussion Meeting on 
Relaxations in Complex Systems, edited by K. L. Ngai, E. 
Riande and G. B. Wright [{\em J. Non-Cryst. Solids} 
{\bf 172-174}, (1994)].
\bibitem{yip}
{\em Transport Theory and Statistical Physics}, edited by S. Yip
and P. Nelson (Marcel Dekker, New York, 1995), Vol.~24, No.~6--8.
\bibitem{kob_rev97}
W. Kob, in {\em Supercooled Liquids: Advances and Novel Applications},
edited by J. T. Fourkas, D. Kivelson, U. Mohanty and K. A. Nelson
(ACS Symposium Series 676, Washington, 1997), pp.~28--44.
\bibitem{kob_rev95}
W. Kob, in {\em Annual Reviews of Computational Physics}, edited by D.
Stauffer (World Scientific, Singapore, 1995), Vol.~3, pp.~1--43.
\bibitem{goetze2}
W. G\"otze and L. Sj\"ogren, {\em Rep. Prog. Phys.} {\bf 55}, 241
(1992).
\bibitem{goetze1}
W. G\"otze, in {\em Liquids, Freezing and the Glass
Transition}, edited by J. P. Hansen, D. Levesque and J. Zinn-Justin
(North-Holland, Amsterdam, 1990), Part~1, pp.~287--503.
\bibitem{goetze_yip}
W. G\"otze and L. Sj\"ogren, in {\em Transport Theory and Statistical 
Physics}, edited by S. Yip and P. Nelson (Marcel Dekker, New York, 1995), 
Vol.~24, No.~6--8, pp.~801--853.
\bibitem{gs_97}
W. G\"otze and L. Sj\"ogren, {\em Phys. Rev. A} {\bf 43}, 5442 (1991).
\bibitem{bl}
J.-L. Barrat and A. Latz, {\em J. Phys.: Condens. Matter} {\bf 2},
4289 (1990).
\bibitem{nk}
M. Nauroth and W. Kob, {\em Phys. Rev. E} {\bf 55}, 657 (1997).
\bibitem{ki-cu}
X. C. Zeng, D. Kivelson and G. Tarjus, {\em Phys. Rev. E} {\bf 50}, 1711
(1994); P. K. Dixon, N. Menon and S. R. Nagel, {\em ibid.} {\bf 50}, 1717 (1994);
H. Z. Cummins and G. Li, {\em ibid.} {\bf 50}, 1720 (1994).
\bibitem{tgms}
T. Franosch, W. G\"otze, M. R. Mayr and A. P. Singh, {\em Phys. Rev. E}
{\bf 55}, 3183 (1997).
\bibitem{ss}
T. Scheidsteger and R. Schilling, {\em Phys. Rev. E} {\bf 56}, 2932
(1997).
\bibitem{fgm}
M. Fuchs, W. G\"otze and M. R. Mayr, {\em Phys. Rev. E}, in press.
\bibitem{ffgms}
T. Franosch, M. Fuchs, W. G\"otze, M. R. Mayr and A. P. Singh, {\em Phys. Rev. E}
{\bf 55}, 7153 (1997).
\bibitem{zrff}
R. Zorn, D. Richter, B. Frick and B. Farago, {\em Physica A} {\bf 201}, 52
(1993).
\bibitem{hkf}
A. Hofmann, F. Kremer and E. W. Fischer, {\em Physica A} {\bf 201}, 106 (1993).
\bibitem{fghl}
M. Fuchs, W. G\"otze, W. Hildebrand and A. Latz, {\em J. Phys.: Condens. Matter} 
{\bf 4}, 7709 (1992).
\bibitem{ems}
H. Eliasson, B.-E. Mellander and L. Sj\"ogren, {\em Mode-Coupling Analysis
of Amorphous PET}, preprint.
\bibitem{sj_1991}
L. Sj\"ogren, {\em J. Phys.: Condens. Matter} {\bf 3}, 5023 (1991).
\bibitem{ion}
I. C. Halalay, {\em J. Phys.: Condens. Matter} {\bf 8}, 6157 (1996).
\bibitem{bbbp}
K. Binder, J. Baschnagel, S. B\"ohmer and W. Paul, {\em Phil. Mag. B} {\bf 77}, 
591 (1998).
\bibitem{rev96}
J. Baschnagel, {\em J. Phys.: Condens. Matter} {\bf 8}, 9599 (1996).
\bibitem{kb_rev93}
W. Paul and J. Baschnagel, in {\em Monte Carlo and Molecular Dynamics
Simulations in Polymer Science}, edited by K. Binder
(Oxford University Press, New York, 1995), pp.~307--355.
\bibitem{exmct}
J. Baschnagel and M. Fuchs {\em J. Phys.: Condens. Matter} {\bf 7}, 6761 (1995).
\bibitem{kb}
W. Kob and J.-L. Barrat, {\em Phys. Rev. Lett.} {\bf 78}, 4581 (1997).
\bibitem{bckm}
J.-P. Bouchaud, L. F. Cugliandolo, J. Kurchan and M. M\'ezard, in {\em
Spin Glasses and Random Fields}, edited by A. Young (World Scientific,
Singapore, 1998), pp.~161--223.
\bibitem{bpbd}
C. Bennemann, W. Paul, K. Binder and B. D\"unweg, {\em Phys.\ Rev.\ E} {\bf 57},
843 (1997).
\bibitem{bpbb_pressure}
C. Bennemann, W. Paul, J. Baschnagel and K. Binder, {\em Investigating the influence of
different thermodynamic paths on the structural relaxation in a glass forming polymer
melt}, submitted to {\em J. Phys.: Condens. Matter}. 
\bibitem{kks}
S. K\"ammerer, W. Kob and R. Schilling, {\em Phys. Rev. E} {\bf 58}, 2131 (1998).
\bibitem{cldhs}
H. Z. Cummins, G. Li, W. Du, Y. H. Hwang and G. Q. Shen, {\em Prog. Theo.
Phys. Suppl.} {\bf 126}, 21 (1997).
\bibitem{ka1}
W. Kob and H. C. Andersen, {\em Phys. Rev. E} {\bf 51}, 4626 (1995).
\bibitem{pbfksf}
W. Petry, E. Bartsch, F. Fujara, M. Kiebel, H. Sillescu and B. Farago,
{\em Z. Phys. B} {\bf 83}, 175 (1991).
\bibitem{ka2}
W. Kob and H. C. Andersen, {\em Phys. Rev. E} {\bf 52}, 4134 (1995).
\bibitem{fhl}
M. Fuchs, I. Hofacker and A. Latz, {\em Phys. Rev. A} {\bf 45}, 898 (1992).
\bibitem{bbpb_msd+rouse}
C. Bennemann, J. Baschnagel, W. Paul and K. Binder, {\em Molecular-Dynamics
Simulation of a Glassy Polymer Melt: Rouse Modes and Mean-Square Displacements},
in preparation.
\bibitem{mf_ali93}
M. Fuchs, {\em J. Non-Cryst. Solids} {\bf 172--174}, 241 (1994).
\bibitem{sfct}
F. Sciortino, L. Fabian, S.-H. Chen and P. Tartaglia, {\em Phys. Rev. E}
{\bf 56}, 5397 (1997).
\bibitem{kks1}
S. K\"ammerer, W. Kob and R. Schilling, {\em Phys. Rev. E} {\bf 56}, 
5450 (1998).
\bibitem{fghl1}
M. Fuchs, W. G\"otze, W. Hildebrand and A. Latz, {\em Z. Phys. B} 
{\bf 87}, 43 (1992).
\bibitem{lw}
L. J. Lewis and G. Wahnstr\"om, {\em Phys. Rev. E} {\bf 50}, 3865 (1994).
\bibitem{cdfghllt}
H. Z. Cummins, W. M. Du, M. Fuchs, W. G\"otze, S. Hildebrand, A. Latz,
G. Li and N. J. Tao, {\em Phys. Rev. E} {\bf 47}, 4223 (1993).
\bibitem{vmu} 
W. van Megen and S. M. Underwood, {\em Phys. Rev. E} {\bf 49}, 4206 (1994).
\bibitem{fcdgllt}
M. Fuchs, H. Z. Cummins, W. M. Du, W. G\"otze, A. Latz,
G. Li and N. J. Tao, {\em Phil. Mag. B} {\bf 71}, 771 (1995).
\bibitem{bjncs_95}
E. Bartsch, {\em J. Non-Cryst. Solids} {\bf 193}, 384 (1995).
\bibitem{jkba}
J. Horbach, W. Kob, K. Binder and C. A. Angell, {\em Phys. Rev. E}
{\bf 54}, R5897 (1996).
\bibitem{zorn97}
R. Zorn, {\em Phys. Rev. B} {\bf 55}, 6249 (1997).
\bibitem{col95}
J. Colmenero, {\em Macromol. Symp.} {\bf 94}, 105 (1995).
\bibitem{col93}
J. Colmenero, {\em Physica A} {\bf 201}, 38 (1993).
\bibitem{bbbp98}
K. Binder, C. Bennemann, J. Baschnagel and W. Paul, {\em Anomalous diffusion
of polymers in supercooled melts near the glass transition}, to be published
by Springer Verlag, Berlin.
\bibitem{bipa97}
K. Binder and W. Paul, {\em J. Polymer Sci. Part B: Polymer Physics}
{\bf 35}, 1 (1997).
\bibitem{doi}
M. Doi and S. F. Edwards, {\em The Theory of Polymer 
Dynamics} (Clarendon Press, Oxford, 1986).
\end{thebibliography}
\end{document}